\documentclass[aps,pre,twocolumn,floats,showpacs]{revtex4}
\usepackage{epsfig}
\usepackage{color}
\usepackage{bm}
\usepackage{latexsym}

\begin{document}
\newcommand{\hide}[1]{}
\newcommand{\tbox}[1]{\mbox{\tiny #1}}
\newcommand{\half}{\mbox{\small $\frac{1}{2}$}}
\newcommand{\sinc}{\mbox{sinc}}
\newcommand{\const}{\mbox{const}}
\newcommand{\trc}{\mbox{trace}}
\newcommand{\intt}{\int\!\!\!\!\int }
\newcommand{\ointt}{\int\!\!\!\!\int\!\!\!\!\!\circ\ }
\newcommand{\eexp}{\mbox{e}^}
\newcommand{\bra}{\left\langle}
\newcommand{\ket}{\right\rangle}
\newcommand{\EPS} {\mbox{\LARGE $\epsilon$}}
\newcommand{\ar}{\mathsf r}
\newcommand{\im}{\mbox{Im}}
\newcommand{\re}{\mbox{Re}}
\newcommand{\bmsf}[1]{\bm{\mathsf{#1}}}
\newcommand{\mpg}[2][1.0\hsize]{\begin{minipage}[b]{#1}{#2}\end{minipage}}

\title{A Lloyd-model generalization: Conductance fluctuations in one-dimensional disordered systems}

\author{J. A. M\'endez-Berm\'udez,$^1$ A. J. Mart\'inez-Mendoza,$^{1,2}$ V. A. Gopar,$^3$ and I. Varga$^2$}
\affiliation{$^1$Instituto de F\'{\i}sica, Benem\'erita Universidad
Aut\'onoma de Puebla, Apartado Postal J-48, Puebla 72570, Mexico \\
$^2$Elm\'eleti Fizika Tansz\'ek, Fizikai Int\'ezet, Budapesti M\H uszaki
\'es Gazdas\'agtudom\'anyi Egyetem, H-1521 Budapest, Hungary \\
$^3$Departamento de F\'isica Te\'orica, Facultad de Ciencias, 
and BIFI, Universidad de Zaragoza, Pedro Cerbuna 12, E-50009, Zaragoza, Spain}

\date{\today}

\begin{abstract}
We perform a detailed numerical study of the conductance $G$ through one-dimensional (1D) 
tight-binding wires with on-site disorder. The random configurations of the on-site energies $\epsilon$ 
of the tight-binding Hamiltonian are characterized by long-tailed distributions: For large 
$\epsilon$, $P(\epsilon)\sim 1/\epsilon^{1+\alpha}$ with $\alpha\in(0,2)$. Our  model serves 
as a generalization of 1D Lloyd's model, which corresponds to $\alpha=1$. First, we verify 
that the ensemble average $\bra -\ln G\ket$ is proportional to the length of the wire $L$ for all values of 
$\alpha$, providing the localization length $\xi$ from $\bra-\ln G\ket=2L/\xi$. Then, 
we show that the probability distribution function $P(G)$ is fully determined by the exponent 
$\alpha$ and $\bra-\ln G\ket$. In contrast to 1D wires with 
standard white-noise disorder, our wire model exhibits bimodal distributions of the conductance 
with peaks at $G=0$ and $1$. In addition, we show that $P(\ln G)$ is proportional to 
$G^\beta$, for $G\to 0$, with $\beta\le\alpha/2$, in agreement to previous studies.
\end{abstract}


\maketitle

\section{Introduction and model}

The recent experimental realizations of the so-called L\'evy glasses \cite{BBW08} as well as 
``L\'evy waveguides'' \cite{FMG14} has refreshed the interest in the study of systems 
characterized by L\'evy-type disorder (see for example 
Refs.~\cite{BGA09,BCV10,EVM10,BVP10,BBV10,BRS12,GAB12,BSLV12,BME13,BUV14,BBL14,ZLW15,AN15}). 
That is, disorder characterized by random variables 
$\{\epsilon\}$ whose density distribution function exhibits a slow
decaying tail:
\begin{equation}
\label{rho}
P(\epsilon) \sim \frac{1}{\epsilon^{1+\alpha}} \ ,
\label{Pofepsilon}
\end{equation}
for large $x$, with $0<\alpha<2$ (this kind of probability distributions are known as 
$\alpha$-stable distributions \cite{uchaikin}). In fact, the study of this class of disordered 
systems dates back to Lloyd \cite{L69}, who  studied spectral properties of a 
three-dimensional (3D) lattice described by a 3D tight-binding Hamiltonian with 
Cauchy-distributed on-site potentials [which corresponds to the particular value $\alpha=1$
in Eq.~(\ref{Pofepsilon})]. Since then, a considerable number of works have been devoted to 
the study of spectral, eigenfunction, and transport properties of Lloyd's model in its original 
3D setup \cite{S70,KB73,H76,BG77,KG79,JK83,S83,RW86,RPW86,KK88}
and  in lower dimensional versions
\cite{JK83,RPW86,T72,I73,AT74,T83,M84,RK86,S86,FPG89,CGIFM92,MBHGZ98,DLA00,GF01,FB01,TS03,RK07,K12}.

Of particular interest is the comparison between the one-dimensional (1D) Anderson model
(1DAM) \cite{A58} and the 1D Lloyd's model, since the former represents the most prominent
model of disordered wires \cite{50years}. Indeed, both models are described by the 1D
tight-binding Hamiltonian:
\begin{eqnarray}
\label{H}
H & = & \sum_{n=1}^L \left[ \epsilon_n \left.\mid n \ket\bra n \mid\right. \right. \nonumber \\
& & \left. - \nu_{n,n+1} \left.\mid n \ket\bra n+1 \mid - \nu_{n,n-1} \mid n \ket\bra n-1 \mid\right. \right] ;
\end{eqnarray}
where $L$ is the length of the wire given as the total number of sites $n$, 
$\epsilon_n$ are random on-site potentials, and $\nu_{n,m}$ are the hopping integrals between
nearest  neighbors (which are set to a constant value $\nu_{n,n\pm 1}=\nu$).
However, while for the standard 1DAM (with white-noise on-site disorder
$\bra\epsilon_n\epsilon_m\ket=\sigma^2\delta_{nm}$ and $\bra\epsilon_n\ket=0$) the on-site potentials
are characterized by a finite variance $\sigma^2=\bra\epsilon_n^2\ket$
(in most cases the corresponding probability distribution function $P(\epsilon)$ is chosen as
a box or a Gaussian distribution), in the Lloyd's model the variance $\sigma^2$ of the random on-site energies
$\epsilon_n$ diverges since they follow a Cauchy distribution. 

It is also  known that the eigenstates $\Psi$ of the {\it infinite} 1DAM are exponentially
localized around a site position $n_0$ \cite{50years}:
\begin{equation}
\label{Psi}
|\Psi_n|\sim\exp\left( -\frac{|n-n_0|}{\xi} \right) \ ;
\end{equation}
where $\xi$ is the eigenfunction localization length. Moreover, for weak disorder ($\sigma^2\ll 1$), the
only relevant parameter for describing the statistical properties of the transmission of the {\it finite} 1DAM is 
the ratio $L/\xi$ \cite{ATA80}, a fact
known as single parameter scaling. The above  exponential localization of eigenfunctions
makes the transmission or dimensionless conductance $G$ 
exponentially small, i.e.,~\cite{LGP88}
\begin{equation}
\bra- \ln G \ket = \frac{2L}{\xi} \ ;
\label{linfeff}
\end{equation}
thus, this relation can be used to obtain 
the localization length. Remarkably, it has been shown that Eq.~(\ref{linfeff}) is
also valid for the 1D Lloyd's model \cite{TS03} implying a single parameter scaling, see 
also \cite{DLA00}.

It is also relevant to mention that studies of transport quantities through 1D wires with
L\'evy-type disorder, different from the 1D Lloyd's model, have been reported.
For example, wires with scatterers randomly spaced along the wire according to a
L\'evy-type distribution were studied in Refs.~\cite{BGA09,BCV10,FG10,FMG12}.
Concerning the conductance of such wires, a prominent result reads that the corresponding
probability distribution function $P(G)$ is fully determined by the exponent $\alpha$ of
the power-law decay of the L\'evy-type distribution and the average (over disorder realizations) 
$\bra -\ln G \ket$ \cite{FG10,FMG12}; i.e.,~all other details of the disorder configuration 
are irrelevant. In this sense, $P(G)$ shows {\it universality}. 
Moreover, this fact was already verified experimentally in microwave random waveguides 
\cite{FMG14} and tested numerically using the tight-binding model of
Eq.~(\ref{H}) with $\epsilon_n=0$ and off-diagonal L\'evy-type disorder \cite{AKFG12}
(i.e.,~with $\nu_{n,m}$ in Eq.~(\ref{H}) distributed according to a L\'evy-type distribution).

It is important to point out that 1D tight-binding wires with power-law distributed
random on-site potentials, characterized by power-laws different from $\alpha=1$ (which
corresponds to the 1D Lloyd's model), have been scarcely studied;
for a prominent exception see \cite{TS03}. Thus, in this paper we undertake this task
and study numerically the conductance though disordered wires defined as a generalization
of the 1D Lloyd's model as follows. We shall study 1D wires described by the Hamiltonian 
of Eq.~(\ref{H}) having constant hopping integrals,
$\nu_{n,n\pm 1}=\nu=1$, and random on-site potentials $\epsilon_n$ which follow 
a L\'evy-type distribution with a long tail, like in Eq.~(\ref{Pofepsilon}) with $0<\alpha<2$. 
We name this setup the 1DAM with L\'evy-type on-site disorder. We note that 
when $\alpha=1$ we recover the 1D  Lloyd's model.

Therefore, in the following section we shall show that (i) the conductance distribution
$P(G)$ is fully determined by the power-law exponent $\alpha$ and the ensemble average
$\bra -\ln G \ket$; (ii) for $\alpha\le 1$ and  $\bra -\ln G \ket\sim 1$,   
bimodal distributions for $P(G)$  with peaks at $G \sim 0$ and $G \sim 1$ are obtained, 
revealing the coexistence of insulating and ballistic
regimes; and (iii) the probability 
distribution $P(\ln G)$ is proportional to $G^\beta$, for vanishing $G$, with 
$\beta\le\alpha/2$.

\section{Results and discussion}

Since we are interested in the conductance statistics of the 1DAM with L\'evy-type
on-site disorder we have to define first the scattering setup we shall use:
We open the isolated samples described above by attaching two semi-infinite
single channel leads to the border sites at opposite sides of the 1D wires.
Each lead is also described by a 1D semi-infinite tight-binding
Hamiltonian. Using the Heidelberg approach \cite{MW69} we can write the 
transmission amplitude through the disordered wires as 
$t = -2i \sin (k)\, {\cal W}^{\,T} (E-{\cal H}_{\rm eff})^{-1} {\cal W}$,
where $k=\arccos(E/2)$ is the wave vector supported in the leads and
${\cal H}_{\rm eff}$ is an effective non-hermitian Hamiltonian given by
${\mathcal{H}}_{\rm eff}=H- e^{ik} {\cal W}{\cal W}^{\,T}$.
Here, ${\cal W}$ is a $L\times 1$ vector that specifies the positions
of the attached leads to the wire. In our setup, all elements of ${\cal W}$
are equal to zero except ${\cal W}_{11}$ and ${\cal W}_{L1}$ which we set to
unity (i.e.,~the leads are attached to the wire with a strength equal to the
inter-site hopping amplitudes: $\nu=1$). Also, we have fixed the energy at $E=0$ in all 
our calculations, although the same conclusions are obtained for $E \ne 0$.
Then, within a scattering approach to the electronic transport, we compute the
dimensionless conductance as \cite{Landauer} $G=|t|^2$.

\begin{figure}[t]
\centerline{\includegraphics[width=7cm]{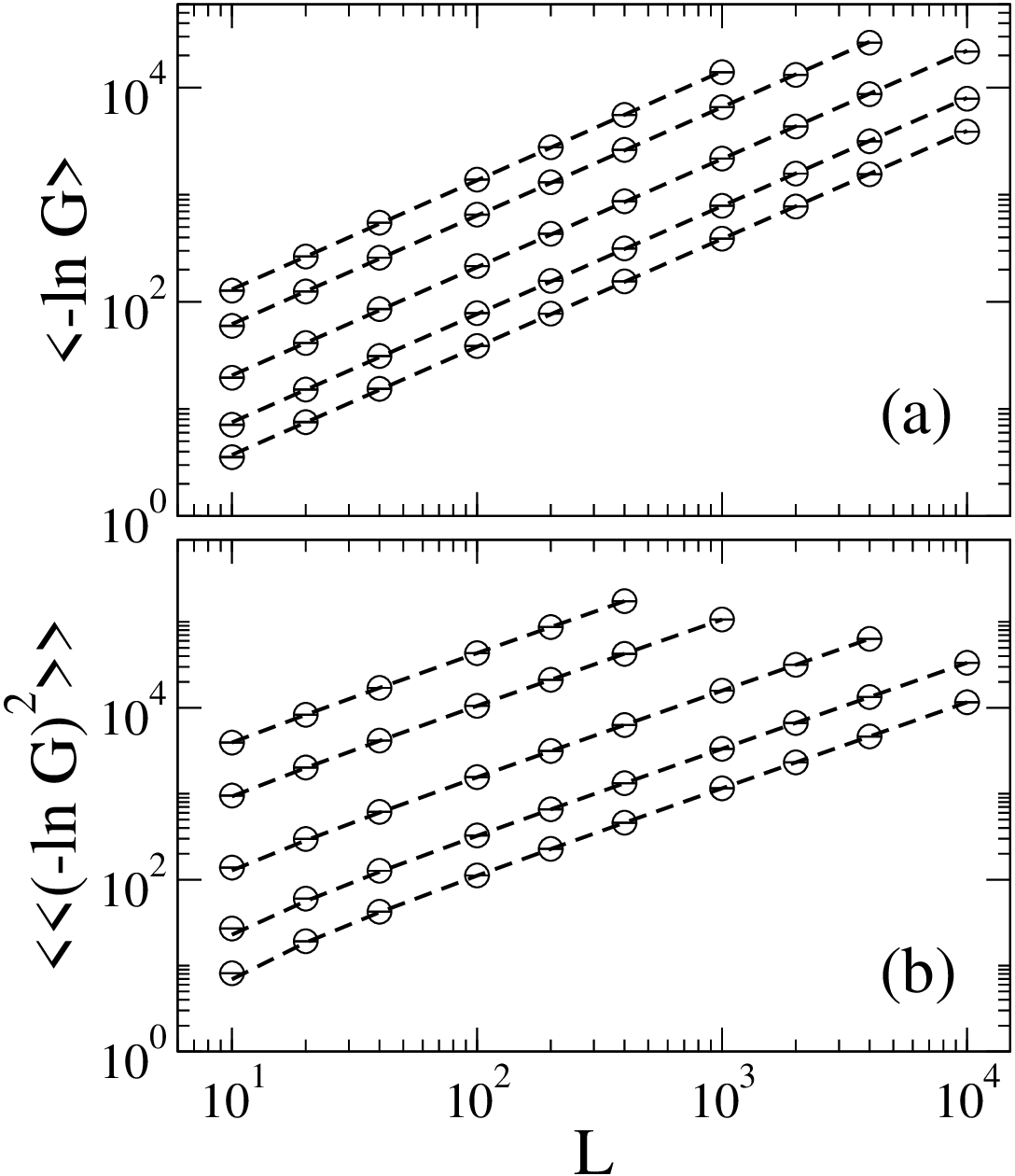}}
\caption{(a) Average logarithm of the conductance $\bra -\ln G \ket$ as a function
of $L$ for the 1DAM with L\'evy-type on-site disorder (symbols).
Dashed lines are the fittings of the data with Eq.~(\ref{linfeff}) used to extract
$\xi$. (b) $\bra\bra (-\ln G)^2 \ket\ket$ as a function of $L$ (symbols). Dashed lines are
fittings of the data with the function $\bra\bra (-\ln G)^2 \ket\ket = 4c_2L$, see Eq.~(\ref{ck}).
In both panels $\alpha = 1/10$, 1/5, 1/2, 1, and 3/2 (from top to bottom).
Each point was calculated using $10^4$ disorder realizations. $E=0$ was used.}
\label{Fig1}
\end{figure}
\begin{figure}[t]
\centerline{\includegraphics[width=8.5cm]{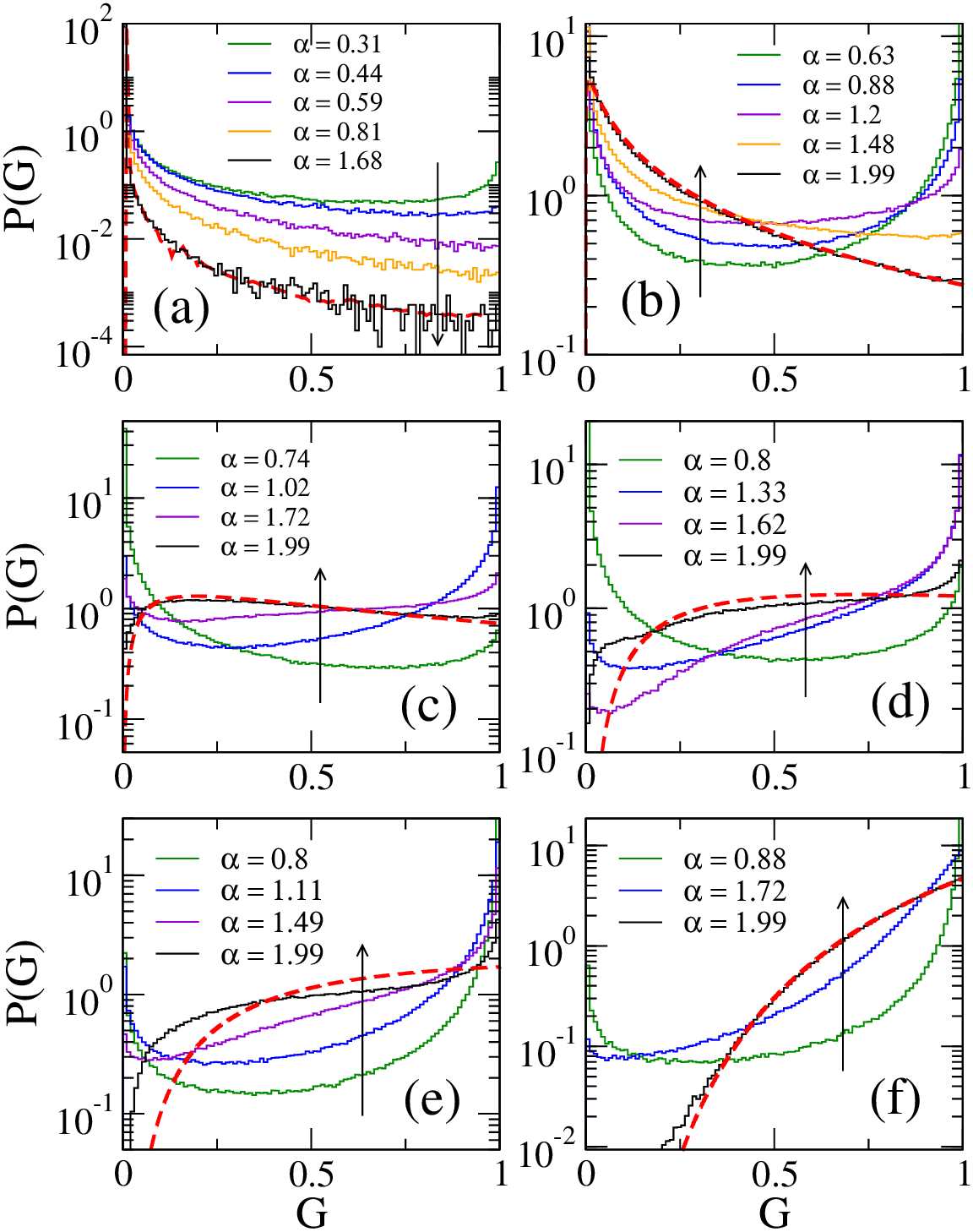}}
\caption{(Color online) Conductance distribution $P(G)$ for the 1DAM with
L\'evy-type disorder (histograms). Each panel correspond to a fixed value of
$\bra -\ln G \ket$: (a) $\bra -\ln G \ket=20$, (b) $\bra -\ln G \ket=2$,
(c) $\bra -\ln G \ket=1$, (d) $\bra -\ln G \ket=2/3$, 
(e) $\bra -\ln G \ket=1/2$, and (f) $\bra -\ln G \ket=1/5$.
In each panel we include histograms for several values of $\alpha$, 
where $\alpha$ increases in the arrow direction. 
$E=0$ was used. Each histogram was calculated using $10^6$
disorder realizations. The red dashed lines are the theoretical predictions
of $P(G)$ for the 1DAM with white noise disorder $P_{\tbox{WN}}(G)$
corresponding to the particular value of $\bra -\ln G \ket$ of each panel.}
\label{Fig2}
\end{figure}
\begin{figure}[t]
\centerline{\includegraphics[width=8.5cm]{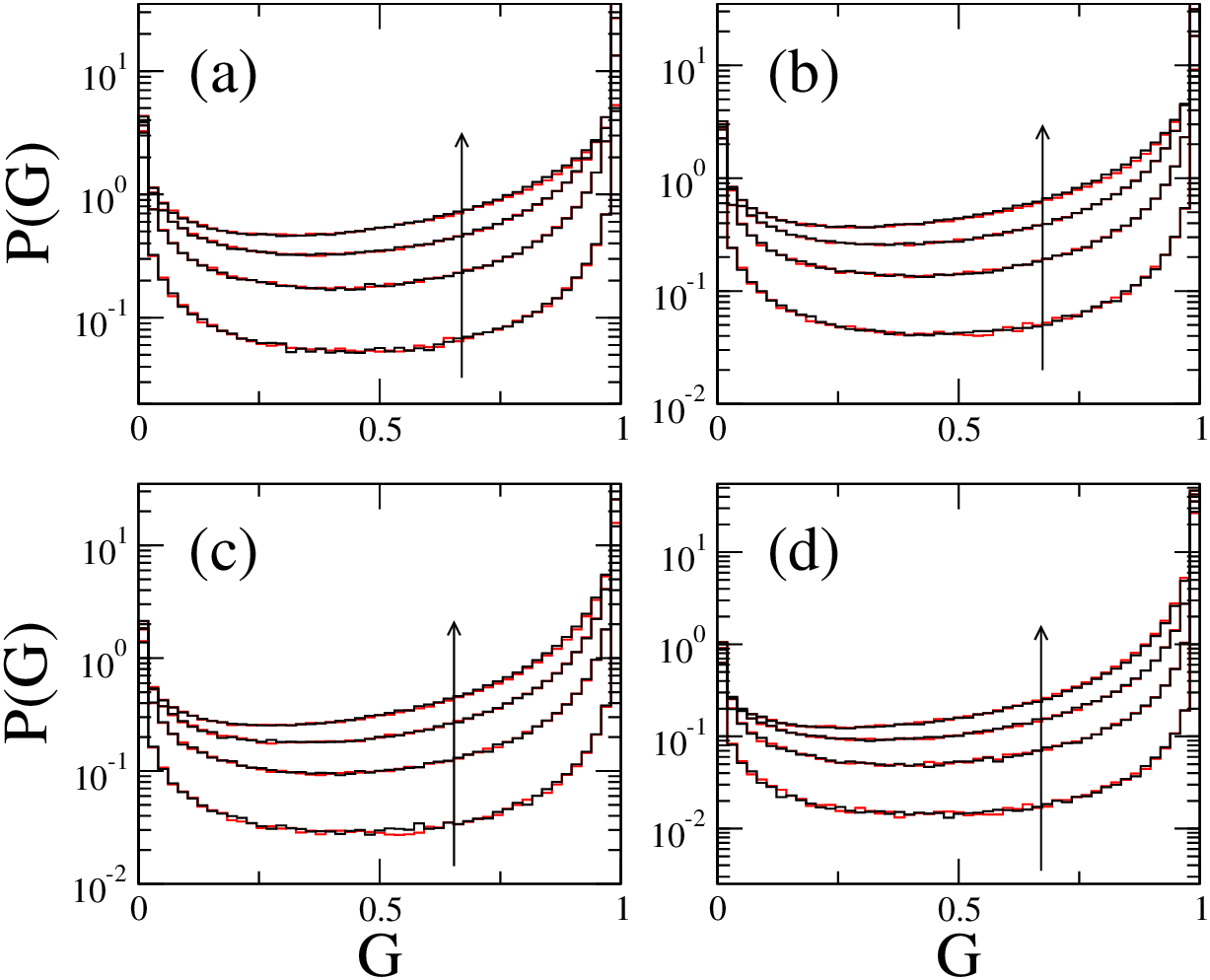}}
\caption{(Color online) Conductance distribution $P(G)$ for the 1DAM with 
L\'evy-type on-site disorder. Each panel corresponds to a fixed value of 
$\bra -\ln G \ket$: (a) $\bra -\ln G \ket=1$, (b) $\bra -\ln G \ket=3/4$,
(c) $\bra -\ln G \ket=1/2$, and (d) $\bra -\ln G \ket=1/4$. 
In each panel we include histograms for $\alpha = 1/4$, 
1/2, 3/4, and 1; where $\alpha$ increases in the arrow direction. $E=0$ was 
used. For each value of $\alpha$ we present two histograms using different 
L\'evy-type density distributions of on-site disorder: $\rho_1(\epsilon)$ in 
red and $\rho_2(\epsilon)$ in black; see Eqs.~(\ref{dd1}) and (\ref{dd2}) in 
\cite{note2}. Each histogram was calculated using $10^6$ disorder realizations.}
\label{Fig3}
\end{figure}

First, we present in Fig.~\ref{Fig1}(a) the ensemble average $\bra -\ln G \ket$
as a function of $L$ for the 1DAM with L\'evy-type disorder for several values of 
$\alpha$. It is clear from this figure that $\bra -\ln G \ket \propto L$ for all
the values of $\alpha$ we consider here. 
Therefore, we can extract the localization length $\xi$ by fitting the curves
$\bra -\ln G \ket$ vs.~$L$ with Eq.~(\ref{linfeff}); see dashed lines in 
Fig.~\ref{Fig1}(a). This behavior should be contrasted to the case of 1D wires with 
off-diagonal L\'evy-type disorder \cite{note0} which shows the dependence
$\bra -\ln G \ket \propto L^{1/2}$ when $\alpha=1/2$ at $E=0$ \cite{AKFG12}.

Also, we have confirmed that the cumulants $\bra\bra (-\ln G)^k \ket\ket$ obey a 
linear relation with the wire length \cite{TS03,TS03b}, i.e.,
\begin{equation}
\lim_{L \to \infty} \frac{\bra\bra (-\ln G)^k \ket\ket}{L} = 2^kc_k \ ,
\label{ck}
\end{equation}
where the coefficients $c_k$, with $c_1\equiv\xi^{-1}$, characterize the Lyapunov
exponent of a generic 1D tight-binding wire with on-site disorder. We have verified 
the above relation, Eq.~(\ref{ck}), for $k=1$, 2, and 3; as an example
in Fig.~\ref{Fig1}(b) we present the results for $\bra\bra (-\ln G)^2 \ket\ket$ as 
a function of $L$ for different values of $\alpha$. The dashed lines are fittings 
of the numerical data (open dots) with the function 
$\bra\bra (-\ln G)^2 \ket\ket = 4c_2L$, see Eq.~(\ref{ck}), which can be used to 
extract the higher order coefficient $c_2$.

Now, in Fig.~\ref{Fig2} we show different conductance distributions $P(G)$ for the 1DAM 
with L\'evy-type on-site disorder for fixed values of $\bra -\ln G \ket$; note that 
fixed $\bra -\ln G \ket$ means fixed ratio $L/\xi$. Several values of $\alpha$ are 
reported in each panel. 
We can observe that for fixed $\bra -\ln G \ket$, by increasing $\alpha$ the 
conductance distribution evolves towards the $P(G)$ corresponding to the 1DAM with 
white noise disorder, $P_{\tbox{WN}}(G)$, as expected. The curves for 
$P_{\tbox{WN}}(G)$ are included as a reference in all panels of Fig.~\ref{Fig2} as 
red dashed lines \cite{note1}. 
In fact, $P(G)$ already corresponds to $P_{\tbox{WN}}(G)$ once $\alpha=2$.

We recall that for 1D tight-binding wires with off-diagonal L\'evy-type disorder  
$P(G)$ is fully determined by the exponent $\alpha$ and the average $\bra -\ln G \ket$ 
\cite{AKFG12}. It is therefore pertinent to ask whether this property 
also holds for {\it diagonal} L\'evy-type disorder. Thus, in Fig.~\ref{Fig3} we show 
 $P(G)$ for the 1DAM with L\'evy-type on-site 
disorder for several values of $\alpha$, where each panel corresponds to a fixed 
value of $\bra -\ln G \ket$.
For each combination of $\bra -\ln G \ket$ and $\alpha$ we present two 
histograms (in red and black) corresponding to wires with on-site random potentials
$\{\epsilon_n\}$ characterized by two {\it different}  density distributions 
\cite{note2}, but with the same exponent $\alpha$ of their corresponding power-law 
tails. We can see from Fig. \ref{Fig3} that for each value of $\alpha$ the 
histograms (in red and black) fall on the top of each other, which is an evidence 
that the conductance distribution 
$P(G)$ for the 1DAM with L\'evy-type on-site disorder is invariant once $\alpha$ 
and $\bra -\ln G \ket$ are fixed; i.e.,~$P(G)$ displays a universal statistics.

Moreover, we want to emphasize the coexistence of insulating and ballistic
regimes characterized, respectively, by the two prominent peaks of $P(G)$ at $G=0$ 
and $G=1$. This behavior, which is more evident for $\bra -\ln G \ket\sim 1$ and 
$\alpha\le 1$ (see Figs.~\ref{Fig2} and \ref{Fig3}), is not observed in 
1D wires with white-noise disorder (see for example the red dashed curves in 
Fig.~\ref{Fig2}). This coexistence of opposite transport regimes has been 
already reported in systems with anomalously localized states: 1D wires with obstacles
randomly spaced according to L\'evy-type density distribution \cite{FG10,AKFG12} 
as well as in  the so-called random-mass Dirac model \cite{SCFG99}.

\begin{figure}[t]
\centerline{\includegraphics[width=8cm]{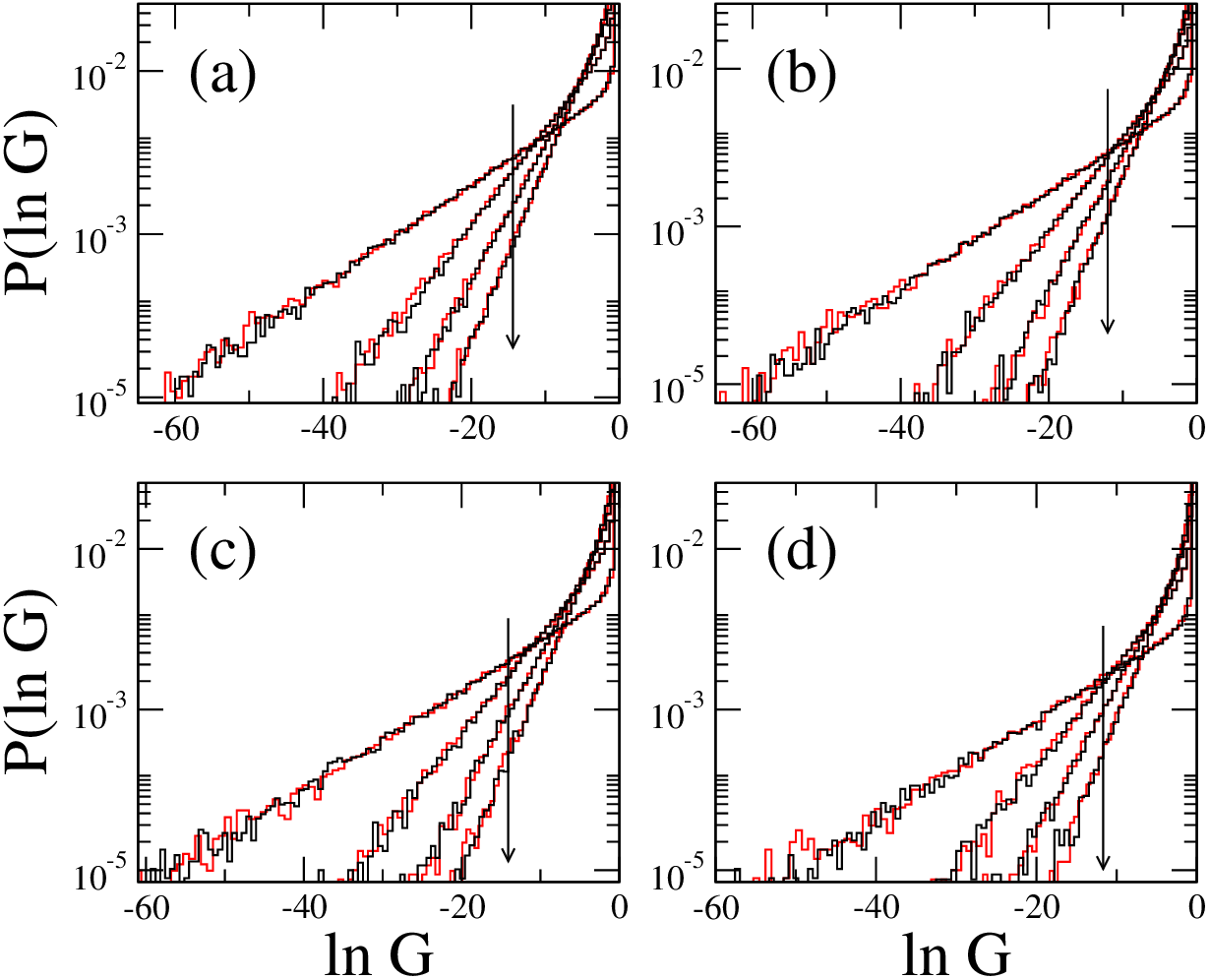}}
\caption{(Color online) Probability distribution functions $P(\ln G)$ for the
1DAM with L\'evy-type on-site disorder. Same parameters as in Fig.~\ref{Fig3}.
We just recall that in each panel we included histograms for $\alpha = 1/4$,
1/2, 3/4, and 1. Here, $\alpha$ increases in the arrow direction.}
\label{Fig4}
\end{figure}

\begin{figure}[t]
\centerline{\includegraphics[width=6cm]{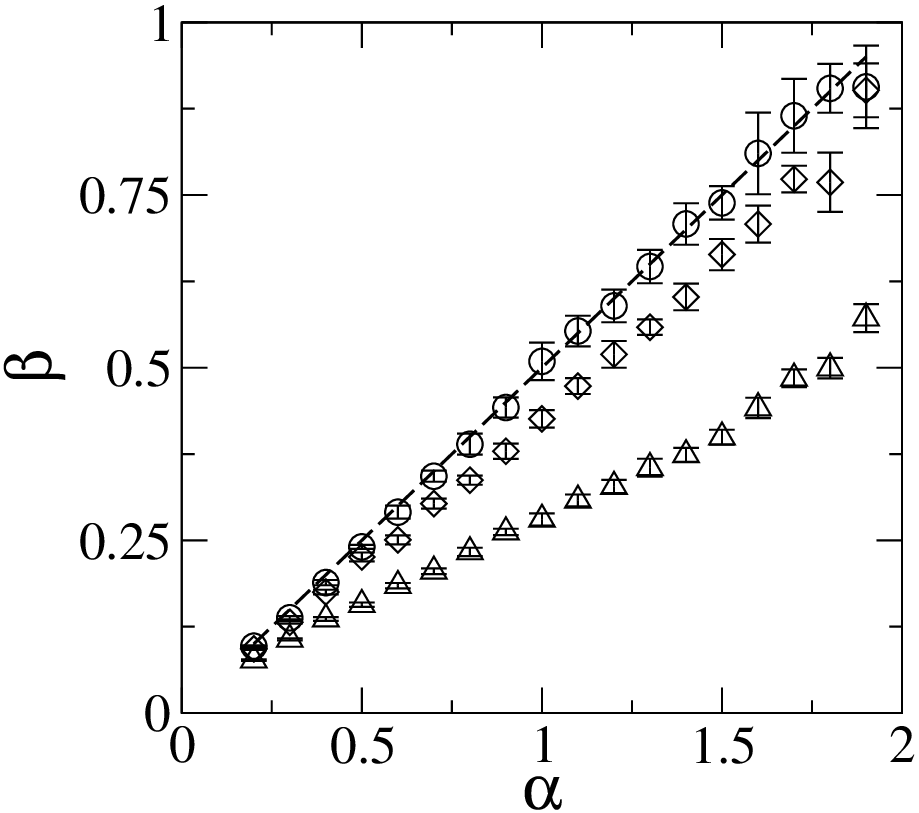}}
\caption{The exponent $\beta$, see Eq.~(\ref{PlnG}), as a function of $\alpha$ for
$\bra-\ln G\ket=1/10$ (circles), 1 (diamonds), and 10 (triangles). The dashed line 
corresponds to $\beta=\alpha/2$.
$\beta$ was obtained from power-law fittings of the tails of the histograms of
$P(\ln G)$ in the interval $P(\ln G)\in [10^{-5},10^{-3}]$.}
\label{Fig5}
\end{figure}

Finally, we study the behavior of the tail of the distribution $P(\ln G)$. 
Thus, using the same data of Fig. \ref{Fig3}, in Fig. \ref{Fig4} we plot 
$P(\ln G)$. As expected, since $P(G)$ is determined by $\alpha$ and 
$\bra -\ln G \ket$, we can see that $P(\ln G)$ is invariant once those 
two quantities ($\alpha$ and $\bra -\ln G \ket$) are fixed (red and 
black histograms fall on top of each other). Moreover, from Fig.~\ref{Fig4}
we can deduce a power-law behavior:
\begin{equation}
P(\ln G) \propto G^\beta
\label{PlnG}
\end{equation}
for $G\to 0$ when $\alpha<2$. For $\alpha=2$, $P(\ln G)$ displays a log-normal
tail (not shown here), expected for 1D systems in the presence of Anderson localization.
Actually, the behavior (\ref{PlnG}) was already anticipated in \cite{TS03} as
$P(G)\sim G^{-(2-\lambda)/2}$ for $G\to 0$ with $\lambda<\alpha$; which in our study
translates as $P(\ln G)\propto G^{\lambda/2}$ (since $P(\ln G)=GP(G)$) with
$\lambda/2\equiv\beta\le\alpha/2$.
Indeed, we have validated the last inequality in Fig.~\ref{Fig5} where
we report the exponent $\beta$ obtained from power-law fittings of
the tails of the histograms of $P(\ln G)$. In addition, we have observed that
the value of $\beta$ depends on the particular value of $\bra -\ln G \ket$ 
characterizing the corresponding histogram of $P(\ln G)$. Also, from Fig.~\ref{Fig5} 
we note that $\beta\approx\alpha/2$ as the value of
$\bra -\ln G \ket$ decreases.


\section{Conclusions}

In this work we have studied the conductance $G$ through a generalization of Lloyd's model in
one dimension: We consider one-dimensional (1D) tight-binding wires with on-site disorder
following a L\'evy-type distribution, see Eq.~(\ref{Pofepsilon}), characterized by the exponent $\alpha$
of the power-law decay. 
We have verified that different cumulants of the variable $\ln G$ decrease linearly with
the length wire $L$. In particular, we were able to extract the eigenfunction
localization length $\xi$ from $\bra-\ln G\ket=2L/\xi$. Then, we have shown some evidence  
that the probability distribution function $P(G)$ is invariant, i.e.,~fully determined, 
once $\alpha$ and $\bra-\ln G\ket$ are fixed; in agreement with other L\'evy-disordered 
wire models \cite{AKFG12,FG10,FMG12,FMG14}.
We have also reported the coexistence of insulating and ballistic  regimes, evidenced 
by peaks in $P(G)$ at $G=0$ and $G=1$; these peaks are most prominent and commensurate 
for $\bra -\ln G \ket\sim 1$ and $\alpha\le 1$.
Additionally we have shown that $P(\ln G)$ develops power-law tails for $G\to 0$, characterized
by the power-law $\beta$ (also invariant for fixed $\alpha$ and $\bra-\ln G\ket$) which,
in turn, is bounded from above by $\alpha/2$.
This upper bound of $\beta$ implies that the smaller the value of $\alpha$ the larger
the probability to find vanishing conductance values in our L\'evy-disordered wires.

\begin{acknowledgments}
J.A.M.-B. and A.J.M.-M. thank F. M. Izrailev and N. M. Makarov for useful comments.
J.A.M.-B. and A.J.M.-M. also thank
FAPESP (Grant No.~2014/25997-0),
CONACyT (Grants No.~I0010-2014-246246 and No.~CB-2013-220624),
VIEP-BUAP (Grant No.~MEBJ-EXC15-I), and
PIFCA (Grant No.~BUAP-CA-169) for financial support.
V.A.G. acknowledges support from MINECO (Spain) under the Project number FIS2012-35719-C02-02. 
\end{acknowledgments}


\begin{thebibliography}{99}
\bibitem{note0}
It is pertinent to remark that the dependence $\bra -\ln G \ket \propto L^{1/2}$, 
when $\alpha=1/2$ at $E=0$, reported in \cite{AKFG12} for 1D wires with 
off-diagonal L\'evy-type disorder was observed when the wire length $L$
was defined as the total sum of the hopping integrals $L=\sum_n\nu_{n,n+1}$.
\bibitem{note1}
Using the results in Ref.~\cite{ioannis}, Eq.~(2) of that work, $P_{\tbox{WN}}(G)$ 
is given by 
\begin{equation}
\label{pofT}
 P_{\tbox{WN}}(G)=C \sqrt{ \frac{\mathrm{acosh} \left( 1/\sqrt{G} \right)}{G^3\sqrt{1-G}}}
\exp\left[ - \frac{1}{s}\mathrm{acosh}^2\left(\frac{1}{\sqrt{G}}\right) \right] ,
\end{equation}
where $C$ is a normalization constant and $s=L/\ell$, $\ell$ being the mean free path. 
The parameter $s$ can the obtained numerically from the ensemble average 
$\langle \ln G \rangle =-L/\ell$.
\bibitem{note2}
We have used the particular density distributions:
\begin{equation}
\rho_1(\epsilon) = \frac{1}{\Gamma(\alpha)}\left(\frac{1}{2}\right)^{\alpha}
\frac{1}{\epsilon^{1+\alpha}} \exp\left(-\frac{1}{2\epsilon}\right)
\label{dd1}
\end{equation}
and
\begin{equation}
\rho_2(\epsilon) = \frac{\alpha}{(1+{\epsilon})^{1+\alpha}} \ ,
\label{dd2}
\end{equation}
where $\Gamma$ is the Euler gamma function.
\end{thebibliography}
\end{document}